\documentclass[aps,pra,twocolumn]{revtex4}

\usepackage{amsfonts}
\usepackage{amsmath}
\usepackage[USenglish]{babel}
\usepackage{graphicx}
\usepackage{color}
\usepackage{amssymb}
\usepackage{esint}
\usepackage{lipsum}

\begin{document}

\title{Magnetoelectric effect in cylindrical topological insulators}

\author{A. Mart\'{i}n-Ruiz}
\email{alberto.martin@nucleares.unam.mx}
\affiliation{Instituto de Ciencia de Materiales de Madrid, CSIC, Cantoblanco, 28049 Madrid, Spain}

\affiliation{Centro de Ciencias de la Complejidad, Universidad Nacional Aut\'{o}noma de M\'{e}xico, 04510 Ciudad de M\'{e}xico, M\'{e}xico}

\begin{abstract}
Topological insulators (TIs) exhibit a quantized magnetoelectric response when time-reversal symmetry is broken on its surface. This unusual electromagnetic (EM) response is a unique macroscopic manifestation of the quantum Hall effect on the TI surface and it is described by a quantized Chern-Simons theory. In this paper, we construct the Green's function (GF) describing the EM response of two topological media separated by a cylindrical interface. This GF, in the appropriate limits, describes the magnetoelectric effect of both i) a cylindrical TI surrounded by a dielectric fluid, and ii) a TI with a cylindrical dielectric-filled cavity. We calculate the EM fields produced by a line charge and a line current near the TI surface and we show that, in addition to the standard image electromagnetic sources, charge and current densities of magnetic monopoles will also appear. We discuss some experimental setups which could be used to test these magnetic monopole fields.
\end{abstract}

\maketitle

\section{Introduction}

The last decade has witnessed an explosion of research in topologically nontrivial states of matter. The best  studied of these are the topological insulators (TIs), which are materials that have an insulating bulk but conducting surface states that are protected against disorder by time-reversal (TR) symmetry \cite{Qi-Review, Hassan-Review}. This astonishing phase of matter has been observed in the HgTe quantum well \cite{Konig}, the $\mbox{Bi} _{1-x} \mbox{Sb} _{x}$ alloy \cite{Hsieh}, and the  stoichiometric crystals $\mbox{Bi} _{2} \mbox{Se} _{3}$, $\mbox{Bi} _{2} \mbox{Te} _{3}$, $\mbox{Sb} _{2} \mbox{Te} _{3}$ and $\mbox{TlBiSe} _{2}$ \cite{Stoichiometric1, Stoichiometric2, Stoichiometric3}. These discoveries encourage to further explore the exotic properties of topological insulators.

Apart from their spectroscopic distinguishing features, topological insulators have unconventional electromagnetic (EM) response characteristics which are unique manifestations of the nontrivial bulk topology on macroscopic scales \cite{Qi-TFT}. The universal part of this unusual EM response may be described by the topological $\theta$ term
\begin{align}
S _{\theta} = \frac{\alpha}{4 \pi ^{2}} \int d ^{4} x \, \theta \, \textbf{E} \cdot \textbf{B} , \label{Action}
\end{align}
where $\textbf{E}$ and $\textbf{B}$ are the electromagnetic fields, $\alpha = e ^{2} / \hbar c$ is the fine structure constant, and $\theta$ is the topological magnetoelectric polarization (axion angle) \cite{Wilczek}. In conventional insulators $\theta = 0$, while for topological insulators $\theta = \pi$, which is the only nonzero value compatible with TR symmetry. Since the integrand in $S _{\theta}$ is a total derivative term, it has no effect on Maxwell's equations in the bulk of the TI. The nontrivial topological property, a surface half-integer quantum Hall effect, manifests only when a TR-breaking perturbation is induced on the surface to gap the surface states, for instance, by means of a thin magnetic coating \cite{Burkov}. In this situation $\theta$ is quantized in odd integer values of $\pi$ such that $\theta = \pm (2n + 1) \pi$, where $2n + 1$ corresponds to the number of Dirac fermions on the surface and the two signs correspond to the two possible orientations of the magnetization in the direction perpendicular to the surface \cite{Qi-TFT}.

The topological response theory defined by the $\theta$ term (\ref{Action}) describes the topological magnetoelectric (TME) effect, where an electric field can induce a magnetic polarization, and a magnetic field can induce an electric polarization. As a result, the magneto-optical Faraday and Kerr rotation \cite{Obukhov, Chang, Maciejko, Tse}, the Casimir effect \cite{MCU-Casimir, Cortijo1, Cortijo2}, as well as the influence of TIs on atomic systems nearby \cite{Song, PRA-Hydrogenlike, EPL-Rydberg, Fang} have been investigated. The most striking consequence of the TME effect, with which we are concerned here, is the image magnetic monopole effect, which consists in the appearance of a magnetic field which resembles that produced by a magnetic monopole when an electric charge is put near to a planar TI surface \cite{Qi-Monopole, Karch, MCU-GreenTI}. While this effect can be in principle measurable, its detection is callenging because a number of screening effects not considered in the theory (\ref{Action}), in addition to the practical difficulty in achieving a full insulating bulk behavior in TIs. A way out of these embarrassing effects and limitations is to replace the point charge by a given distribution of charges and/or currents near the TI surface, as well as using TIs with more intricate geometries. This encourage the development of a general formalism describing the TME effect of spherical and cylindrical TIs in the presence of arbitrary distribution of sources.

The importance of Green's function (GF) techniques is apparent in many branches of physics; its applications ranges from quantum field theory and condensed matter physics to electromagnetism. In the context of the topological field theory defined by Eq. (\ref{Action}), in Ref. \cite{MCU-GreenTI} we initiated a method to calculate the electromagnetic response of planar and spherical TIs by means of Green's function techniques. The dyadic GF for a stack of layered planar TIs has also been constructed in Ref. \cite{Crosse}. From a practical point of view, cylindrical TIs are also amenable for possible experiments designed to test the TME effect. Therefore, the corresponding description of a cylindrical topologically ponderable media is needed. In this work we aim  to fill in this gap by constructing the static Green's function which describes the TME effect of i) a cylindrical TI surrounded by a dielectric fluid and ii) a TI with a cylindrical dielectric-filled cavity. With the use of our GF matrix, we calculate the EM fields produced by a line charge and a line current in parallel with the TI surface and we interpret them in terms of charge and current densities of magnetic monopoles. This can be thought as an extension of the image magnetic monopole effect first derived in Ref. \cite{Qi-Monopole}. Here, we discuss some experimental setups which could be used to test the magnetic monopole fields. For example, when a line charge is put near to the TI surface, the magnetic field of the image wires carrying magnetic charge densities will modify the magnetization of the thin magnetic coating at the TI surface, and such pattern could be in principle detected by magnetic force microscopy. If we replace the line charge by a line current, we find that the image wires carrying magnetic charge currents exert a Lorentz force upon the original wire which can be easily distinguished from that between the line current and its usual electric images. This anomalous force will deflect the trajectory of an electron beam (coming from low-energy electron gun, for instance) drifting above the TI surface, and this deflection should be traced by angle-resolved measurement. In general, the method here elaborated is more general that the method of images which has been often employed in most of the previous works. Therefore, our results provide a precise starting point from where either analytical or numerical calculations can be performed to obtain the EM fields produced by arbitrary sources.

The paper is organized as follows: In the next Section we construct the GF for two semi-infinite ponderable topological media separated by a cylindrical interface. We relegate the derivation of the reduced free GF to the Appendix. Section \ref{Application} is devoted to some simple applications: a line charge and a line current near the TI surface. In Sec. \ref{Discussion} we discuss our results and we propose some possible experimental setups which could be used to test them. Throughout the paper we use Gaussian units and the metric signature will be taken as $(+,-,-,-)$.

\section{Green's function method} \label{GreenSection}

\begin{figure}[t]
\begin{center}
\includegraphics{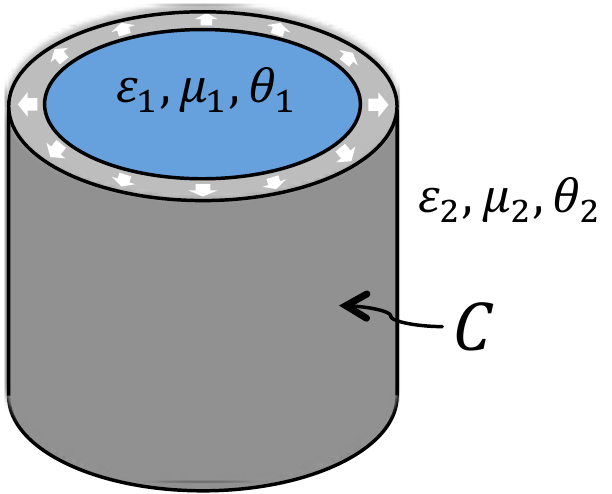}
\end{center}
\caption{{\protect\small Illustration of the theoretical setting used to construct the Green's function. The inner cylinder, of radius $a$, is occupied by a topological medium with dielectric constant $\epsilon _{1}$, magnetic permeability $\mu _{1}$ and axion angle $\theta _{1}$. The exterior region is occupied by a medium with dielectric constant $\epsilon _{2}$, magnetic permeability $\mu _{2}$ and axion angle $\theta _{2}$. The interface is covered with a magnetic layer of small tickness (not to scale).}}
\label{CylindricalTI}
\end{figure}

The electromagnetic response of a system in the presence of the $\theta$ term (\ref{Action}) is still described by the ordinary Maxwell's equations but with the modified constituent equations describing the TME effect \cite{Qi-TFT}
\begin{align}
\textbf{D} = \epsilon \textbf{E} + \alpha (\theta / \pi) \textbf{B} , \qquad \textbf{H} = \textbf{B} / \mu - \alpha (\theta / \pi) \textbf{E} , \label{ConstRel}
\end{align}
where $\epsilon$ and $\mu$ are the permittivity and permeability functions, respectively. Now consider the geometry as shown in Fig. \ref{CylindricalTI}. For the sake of gaining generality, our starting point is the situation in which two topological media are separated by a cylindrical surface of radius $a$, which is assumed to be coated with a thin magnetic layer to gap the surface states such that the theory (\ref{Action}) is valid. At the end of the calculations we could take two limits: (i) $\theta _{2} = 0$ to describe the EM response of a cylindrical TI surrounded by a dielectric fluid and (ii) $\theta _{1} = 0$ to describe the EM response of a TI with a cylindrical dielectric-filled cavity. Now we make some simplifying yet realistic assumptions. Without loss of generality, we henceforth assume $\mu _{1} = \mu _{2} = 1$, which is appropriate for both trivial and topological insulators; and we also consider the surface magnetization always pointing outward the TI bulk [e.g. the direction of the surface magnetization shown in Fig. \ref{CylindricalTI} corresponds to the case (i). For the case (ii) the magnetization is inverted]. Besides, the dependence upon position of the functions $\epsilon$ and $\theta$ is limited to a finite discontinuity across the cylindrical surface $\rho = a$, to be called simply $C$. 

To proceed further we write Maxwell's equations in terms of the electromagnetic potential $A ^{\mu} = (\phi , \textbf{A})$. In the Coulomb gauge the EM potentials satisfy
\begin{align}
- \nabla \cdot \left[ \epsilon (\textbf{r}) \nabla \phi \right] + \frac{\alpha}{\pi} \nabla \theta (\textbf{r}) \cdot \nabla \times \textbf{A} &= 4 \pi \rho , \label{EqE} \\ - \nabla ^{2} \textbf{A} + \frac{\alpha}{\pi} \nabla \theta (\textbf{r}) \times \nabla \phi &= \frac{4 \pi}{c} \textbf{J} . \label{EqB}
\end{align}
We observe that these equations depend only on the space gradient of the axion field. Since $\theta (\textbf{r})$ is a piecewise constant function then $\nabla \theta (\textbf{r}) = (\theta _{2} - \theta _{1}) \delta (C) \hat{\textbf{e}} _{\rho}$, where $\hat{\textbf{e}} _{\rho}$ is the outward unit normal to $C$. Therefore the second term in the left hand side of Eq. (\ref{EqB}) corresponds to a surface current density $\textbf{J} = \sigma _{xy} \delta (C) \hat{\textbf{e}} _{\rho} \times \textbf{E}$, which is induced by the in-plane component of the electric field and is perpendicular to it. This current is nothing but the Hall current with quantized conductivity $\sigma _{xy} = \left( n + 1/2\right) \frac{e ^{2}}{h}$ \cite{Qi-TFT}. This is a manifestation of the fact that a 3D TI described by a $\theta$ term is a dielectric with a quantized Hall effect at the surface. Now we rewrite the field equations in the matrix form 
\begin{equation}
\left[ \mathcal{O} ^{\mu} _{\phantom{\mu} \nu} \right] _{\textbf{r}}  A ^{\nu} = \frac{4 \pi}{c} J ^{\mu} , \label{FieldEqsCyl}
\end{equation}
where $J ^{\mu} = (\rho c , \textbf{J})$ and the differential operator $\left[ \mathcal{O} ^{\mu} _{\phantom{\mu} \nu} \right] _{\textbf{r}}$ reads directly from Eqs. (\ref{EqE}) and (\ref{EqB}). In the cylindrical coordinates $(\rho , \varphi , z)$, with the $z$-axis defined as the symmetry axis of the inner cylinder in Fig. \ref{CylindricalTI}, the differential operator becomes
\begin{widetext}
\begin{align}
\left[ \mathcal{O} ^{\mu} _{\phantom{\mu} \nu} \right] _{\textbf{r}} = \left[ \begin{array}{cccc} \hat{\mathcal{O}} _{\epsilon} & + \tilde{\alpha} \, \delta (C) \sin \varphi \frac{\partial}{\partial z} & - \tilde{\alpha} \, \delta (C) \cos \varphi \frac{\partial}{\partial z} & \frac{\tilde{\alpha}}{a} \, \delta (C) \frac{\partial}{\partial \varphi} \\ + \tilde{\alpha} \, \delta (C) \sin \varphi \frac{\partial}{\partial z} & \hat{\mathcal{O}} & 0 & 0 \\ - \tilde{\alpha} \, \delta (C) \cos \varphi \frac{\partial}{\partial z} & 0 & \hat{\mathcal{O}} & 0 \\ \frac{\tilde{\alpha}}{a} \, \delta (C) \frac{\partial}{\partial \varphi} & 0 & 0 & \hat{\mathcal{O}} \end{array} \right] ,  \label{O-operator-Cyl}
\end{align}
\end{widetext} 
where $\tilde{\alpha} = \alpha \left( \theta _{2} - \theta _{1} \right) / \pi$, 
\begin{align}
\hat{\mathcal{O}} _{\epsilon} = - \nabla \cdot \left[ \epsilon (\rho) \nabla \right] = - \epsilon ( \rho ) \nabla ^{2} - \frac{\partial \epsilon (\rho)}{\partial \rho} \frac{\partial}{\partial \rho} , \label{OperatorO}
\end{align}
and $\hat{\mathcal{O}} = - \nabla ^{2}$. To obtain a general solution for the EM four-potential $A ^{\mu}$ in the presence of arbitrary external sources $J ^{\mu}$, we
introduce the Green's function $G ^{\mu} _{\phantom{\mu} \nu} (\textbf{r} , \textbf{r} ^{\prime})$ solving Eq. (\ref{FieldEqsCyl}) for a pointlike source,
\begin{align}
\left[ \mathcal{O} ^{\mu} _{\phantom{\mu} \nu} \right] _{\textbf{r}} G ^{\nu} _{\phantom{\nu} \sigma} (\textbf{r} , \textbf{r} ^{\prime}) = 4 \pi \eta ^{\mu} _{\phantom{\mu} \sigma} \delta (\textbf{r} - \textbf{r} ^{\prime}) , \label{GFEq}
\end{align}
in such a way that the general solution for the EM four-potential is
\begin{align}
A ^{\mu} (\textbf{r}) = \frac{1}{c} \int G ^{\mu} _{\phantom{\mu} \nu} (\textbf{r} , \textbf{r} ^{\prime}) J ^{\nu} (\textbf{r} ^{\prime}) d ^{3} \textbf{r} ^{\prime} . \label{EMPotSol}
\end{align}
In the following, we discuss the general solution to Eq. (\ref{GFEq}). The GF we consider has translational invariance in the $z$ direction, while this invariance is broken in the $\rho$ direction. In accordance with this symmetry, we start by writing the solution of Eq. (\ref{GFEq}) as
\begin{align}
G ^{\mu} _{\phantom{\mu}\nu} (\textbf{r},\textbf{r} ^{\prime}) & = 4 \pi \int _{- \infty}^{+ \infty} \frac{dk}{2 \pi} e ^{i k ( z - z ^{\prime})} \frac{1}{2 \pi} \sum _{m = - \infty} ^{+ \infty} \notag \\ & \phantom{=} \times \sum _{m ^{\prime} = - \infty}^{+ \infty} g _{m m ^{\prime} , \nu} ^{\mu} \left( \rho , \rho ^{\prime} ; k \right) e ^{i ( m \varphi   - m ^{\prime} \varphi ^{\prime} ) },\label{GF-C} 
\end{align}
where $k$ is the momentum in the $z$ direction. In the following, we suppress the dependence on $k$ of the reduced GF $g _{m m ^{\prime} , \nu} ^{\mu}$. To go forward, we have to derive the differential equation for the reduced Green's function. To this end, we insert the $2+1$ representation of Eq. (\ref{GF-C}) and the completeness relation
\begin{align}
& \delta \left( \varphi  - \varphi ^{\prime} \right) \delta \left( z - z ^{\prime} \right) \notag \\ & = \int _{- \infty} ^{ + \infty} \frac{dk}{2 \pi} e ^{i k ( z-z ^{\prime} )} \frac{1}{2 \pi} \sum _{m , m ^{\prime}} \delta _{m m ^{\prime}} e ^{i ( m\varphi - m ^{\prime} \varphi ^{\prime} )} ,  \label{C-Comp}
\end{align}
into Eq. (\ref{GFEq}) to obtain
\begin{align}
\sum _{m} \left[ \mathcal{O} ^{\mu} _{\phantom{\mu} \nu} \right] _{\rho , m} g _{m m ^{\prime} , \sigma} ^{\nu} \left( \rho , \rho ^{\prime} \right) e ^{im \varphi} \notag \\ = \eta ^{\mu} _{\phantom{\mu}\sigma} \frac{\delta (\rho - \rho ^{\prime})}{\rho} \sum _{m} \delta _{m m^{\prime}} e ^{im \varphi} , \label{Intermediate}
\end{align}
where we have used the linear independence of $e ^{- i m ^{\prime} \varphi ^{\prime}}$ and $e ^{-i k z ^{\prime}}$. Here, $\left[ \mathcal{O} ^{\mu} _{\phantom{\mu} \nu} \right] _{\rho , m}$ is the differential operator defined in Eq. (\ref{O-operator-Cyl}) with the replacements $\partial _{z} \rightarrow i k$ and $\partial _{\varphi} \rightarrow im$. We next multiply Eq. (\ref{Intermediate}) to the left by $e ^{-im ^{\prime \prime} \varphi}$ and integrate with respect to $\varphi$. Using the orthogonality relation $\delta _{mm ^{\prime}} = \frac{1}{2 \pi} \int _{0} ^{2 \pi} e ^{i(m-m ^{\prime}) \varphi} d \varphi$ and defining
\begin{align}
\left[ \mathcal{P} ^{\mu} _{\phantom{\mu} \nu} \right] _{\rho , m m ^{\prime \prime}} = \frac{1}{2 \pi} \int _{0} ^{2 \pi} e ^{im ^{\prime \prime} \varphi} \left[ \mathcal{O} ^{\mu} _{\phantom{\mu} \nu} \right] _{\rho , m} e ^{-im \varphi} d \varphi , \label{P-Operator}
\end{align}
Eq. (\ref{Intermediate}) simplifies to
\begin{align}
\sum _{m ^{\prime \prime}} \left[ \mathcal{P} ^{\mu} _{\phantom{\mu} \nu} \right] _{\rho , m m ^{\prime \prime}} g _{m ^{\prime \prime} m ^{\prime} , \sigma} ^{\nu} \left( \rho , \rho ^{\prime} \right) = \eta ^{\mu} _{\phantom{\mu}\sigma} \frac{\delta (\rho - \rho ^{\prime})}{\rho} \delta _{m m ^{\prime}} . \label{RedGreenCyl-DiffEq}
\end{align}
The differential operator (\ref{P-Operator}) can be explicitly written in the matrix form
\begin{widetext}
\begin{align}
\left[ \mathcal{P} ^{\mu} _{\phantom{\mu} \nu} \right] _{\rho , m m ^{\prime \prime}} = \hspace{0.5cm} \left[ \begin{array}{cccc} \delta _{mm ^{\prime \prime}} \hat{\mathcal{O}} _{\epsilon} & + ik \tilde{\alpha} \, \delta (C) \left< \sin \varphi \right> _{mm ^{\prime \prime}} & - ik \tilde{\alpha} \, \delta (C) \left< \cos \varphi \right> _{mm ^{\prime \prime}} & i m \frac{\tilde{\alpha}}{a} \, \delta (C) \delta _{mm ^{\prime \prime}} \\ + ik \tilde{\alpha} \, \delta (C) \left< \sin \varphi \right> _{mm ^{\prime \prime}} & \delta _{mm ^{\prime \prime}} \hat{\mathcal{O}} & 0 & 0 \\ - ik \tilde{\alpha} \, \delta (C) \left< \cos \varphi \right> _{mm ^{\prime \prime}} & 0 & \delta _{mm ^{\prime \prime}} \hat{\mathcal{O}} & 0 \\ i m \frac{\tilde{\alpha}}{a} \, \delta (C) \delta _{mm ^{\prime \prime}} & 0 & 0 & \delta _{mm ^{\prime \prime}} \hat{\mathcal{O}} \end{array} \right] , \label{P-operator-Cyl}
\end{align}
\end{widetext}
where
\begin{align}
\left< \sin \varphi \right> _{mm ^{\prime \prime}} &= \frac{1}{2i} \left( \delta _{m ^{\prime \prime} , m+1} - \delta _{m ^{\prime \prime} , m-1} \right) , \notag \\ \left< \cos \varphi \right> _{mm ^{\prime \prime}} &= \frac{1}{2} \left( \delta _{m ^{\prime \prime} , m+1} + \delta _{m ^{\prime \prime} , m-1} \right) . \label{integrals}
\end{align}
The solution to Eq. (\ref{RedGreenCyl-DiffEq}) is simple but not straightforward. Here we solve it along the same lines introduced in Refs. \cite{MCU-GreenTI, MCU1, MCU2}. To start, we recall the reduced free Green's function $\mathfrak{g} _{m} (\rho, \rho ^{\prime} )$, satisfying
\begin{align}
\hat{\mathcal{O}} _{\epsilon} \mathfrak{g} _{m} (\rho, \rho ^{\prime}) &= \frac{\delta (\rho - \rho ^{\prime})}{\rho} , \label{DielectricCylRedGreen}
\end{align}
where $\hat{\mathcal{O}} _{\epsilon} = - \rho ^{-1} \partial _{\rho} \left( \epsilon \rho \partial _{\rho} \right) + m ^{2} \rho ^{-2} + k ^{2}$. The reduced free GF $\mathfrak{g} _{m} (\rho, \rho ^{\prime})$ contains all the information concerning the EM response of two dielectrics separated by a cylindrical interface. For the sake of completeness, in the Appendix we sketch the solution to Eq. (\ref{DielectricCylRedGreen}), which we take for granted here. We also define the reduced free-space GF $g _{m} (\rho, \rho ^{\prime})$, which satisfies Eq. (\ref{DielectricCylRedGreen}) for $\epsilon _{1} = \epsilon _{2} = 1$, i.e. $g _{m} (\rho, \rho ^{\prime}) = \lim _{\epsilon _{1} , \epsilon _{2} \rightarrow 1} \mathfrak{g} _{m} (\rho, \rho ^{\prime})$.

Now, in order to solve for the various components of the reduced GF, we observe that the corresponding differential equations fall into two groups, defined by $\sigma = 0$ and $\sigma = i$ in Eq. (\ref{RedGreenCyl-DiffEq}), with $i = 1,2,3$. We first consider the group of equations with $\sigma = 0$:
\begin{align}
\hat{\mathcal{O}} _{\epsilon} g _{m m ^{\prime} , 0} ^{0} + i \frac{\tilde{\alpha}}{a} \delta (C) \sum _{m ^{\prime \prime}} \left< v _{i} \right> _{m m ^{\prime \prime}} g ^{i} _{m ^{\prime \prime} m ^{\prime} , 0} &= \frac{\delta (\rho - \rho ^{\prime})}{\rho} \delta _{mm^{\prime}} , \label{g00-Cyl} \\ \hat{\mathcal{O}} g ^{i} _{m m ^{\prime} , 0} - i \frac{\tilde{\alpha}}{a} \delta (C) \sum _{m ^{\prime \prime} } \left< v ^{i} \right> _{m m ^{\prime \prime}} g _{m ^{\prime \prime} m ^{\prime} , 0} ^{0} &= 0 , \label{gi0-Cyl}
\end{align}
where we have defined the dimensionless vector $\textbf{v} \equiv v ^{i} \hat{\textbf{e}} _{i} = k a \hat{\textbf{e}} _{\varphi} - m \hat{\textbf{e}} _{z} = - k a \sin \varphi \, \hat{\textbf{e}} _{x} + k a \cos \varphi \, \hat{\textbf{e}} _{y} - m \, \hat{\textbf{e}} _{z} $. Equations (\ref{g00-Cyl}) and (\ref{gi0-Cyl}) can be directly integrated by using the reduced free GFs $\mathfrak{g} _{m} (\rho, \rho ^{\prime})$ and $g _{m} (\rho, \rho ^{\prime})$ together with the properties of the Dirac delta function, thus reducing the problem to a set of coupled algebraic equations,
\begin{align}
g _{m m ^{\prime} , 0} ^{0} (\rho , \rho ^{\prime}) &= \delta _{m m ^{\prime}} \mathfrak{g} _{m} (\rho , \rho ^{\prime}) - i \tilde{\alpha} \mathfrak{g} _{m} (\rho , a) \notag \\ & \phantom{=} \times \sum _{m ^{\prime \prime}} \left< v _{i} \right> _{m m ^{\prime \prime}} g ^{i} _{m ^{\prime \prime} m ^{\prime} , 0} (a , \rho ^{\prime}) , \label{g00-Cyl2} \\ g ^{i} _{m m ^{\prime} , 0} (\rho , \rho ^{\prime}) &= i  \tilde{\alpha} g _{m} (\rho , a) \sum _{m ^{\prime \prime} } \left< v ^{i} \right> _{m m ^{\prime \prime}} g _{m ^{\prime \prime} m ^{\prime} , 0} ^{0} (a , \rho ^{\prime}) . \label{gi0-Cyl2}
\end{align}
Now we set $\rho = a$ in Eq. (\ref{gi0-Cyl2}) and then substitute into
Eq. (\ref{g00-Cyl2}), yielding 
\begin{align}
g _{m m ^{\prime} , 0} ^{0} (\rho , \rho ^{\prime}) &= \delta _{mm ^{\prime}} \mathfrak{g} _{m} (\rho , \rho ^{\prime}) - \tilde{\alpha} ^{2} \mathfrak{f} _{m} (k) \notag \\ & \phantom{=} \times \mathfrak{g} _{m} (\rho , a) g _{m m ^{\prime} , 0} ^{0} (a , \rho ^{\prime}) , \label{g00-Cyl4}
\end{align}
where
\begin{align}
\mathfrak{f} _{m} (k) = m ^{2} g _{m}(a,a) + \frac{k ^{2} a ^{2}}{2} \left[ g _{m+1}(a,a) + g _{m-1}(a,a) \right] . \label{f-function-Cyl}
\end{align}
In deriving Eq.~(\ref{g00-Cyl4}) we have used the result $\mathfrak{f} _{m} (k) \delta _{mm ^{\prime \prime \prime}} = - \sum _{m ^{\prime \prime}} g _{m ^{\prime \prime}} (a,a) \left< v _{i} \right> _{m m ^{\prime \prime}} \left< v ^{i} \right> _{m ^{\prime \prime} m ^{\prime \prime \prime}} $, which can be verified directly from Eq. (\ref{integrals}). Solving for $g _{m m ^{\prime} , 0} ^{0} (a , \rho ^{\prime})$ by setting $\rho = a$ in Eq. (\ref{g00-Cyl4}) and inserting the result back into that equation, we finally obtain
\begin{align}
g _{m m ^{\prime} , 0} ^{0} (\rho , \rho ^{\prime}) & = \delta _{mm ^{\prime}} \Big[ \mathfrak{g} _{m} (\rho , \rho ^{\prime}) - \frac{\tilde{\alpha} ^{2} \mathfrak{f} _{m} (k) }{1 + \tilde{\alpha} ^{2} \mathfrak{f} _{m} (k) \mathfrak{g} _{m} (a,a)} \notag \\ & \phantom{=} \times \mathfrak{g} _{m} (\rho , a) \mathfrak{g} _{m} (a , \rho ^{\prime}) \Big] . \label{g00-CylFIN}
\end{align}
The remaining components can be computed directly by substituting $g _{m m ^{\prime} , 0} ^{0} (a , \rho ^{\prime})$ in Eq. (\ref{gi0-Cyl2}). The result is
\begin{align}
g _{m m ^{\prime} , 0} ^{i} (\rho , \rho ^{\prime}) &= \frac{i \tilde{\alpha} \left< v ^{i} \right> _{m m ^{\prime}}}{1 + \tilde{\alpha} ^{2} \mathfrak{f} _{m} (k) \mathfrak{g} _{m} (a,a)} g _{m} (\rho , a ) \mathfrak{g} _{m ^{\prime}} ( a , \rho ^{\prime} ) . \label{gi0-CylFIN}
\end{align}
The second group of equations, defined by $\sigma = i$ in Eq. (\ref{RedGreenCyl-DiffEq}), is
\begin{align}
\hat{\mathcal{O}} _{\epsilon} g ^{0} _{m m ^{\prime} , j} + i \frac{\tilde{\alpha}}{a} \delta (C) \sum _{m ^{\prime \prime}} \left< v _{i} \right> _{m m ^{\prime \prime}} g ^{i} _{m ^{\prime \prime} m ^{\prime} , j} = 0 , \label{g0j-Cyl} \\  \hat{\mathcal{O}} g ^{i} _{m m ^{\prime} , j} - i \frac{\tilde{\alpha}}{a} \delta (C) \sum _{m ^{\prime \prime}} \left< v ^{i} \right> _{m m ^{\prime \prime}} g ^{0} _{m ^{\prime \prime} m ^{\prime} , j} \notag \\ = \eta ^{i} _{\phantom{i}j} \delta _{mm ^{\prime}} \frac{\delta (\rho - \rho ^{\prime})}{\rho} , \label{gij-Cyl}
\end{align}
where $i ,j = 1,2,3$. Integrating these equations we obtain
\begin{align}
g ^{0} _{m m ^{\prime} , j} (\rho , \rho ^{\prime}) &= i \tilde{\alpha} \mathfrak{g} _{m} (\rho , a) \sum _{m ^{\prime \prime}} \left< v _{i} \right> _{m m ^{\prime \prime}} g ^{i} _{m ^{\prime \prime} m ^{\prime} , j} (a, \rho ^{\prime}) , \label{g0j-Cyl2} \\ g ^{i} _{m m ^{\prime} , j} (\rho , \rho ^{\prime}) &= \eta ^{i} _{\phantom{i}j} \delta _{mm ^{\prime}} g _{m} (\rho , \rho ^{\prime}) + i \tilde{\alpha} g _{m} (\rho , a) \notag \\ & \phantom{=} \times \sum _{m ^{\prime \prime}} \left< v ^{i} \right> _{m m ^{\prime \prime}} g ^{0} _{m ^{\prime \prime} m ^{\prime} , j} (a , \rho ^{\prime}) . \label{gij-Cyl2}
\end{align}
Now we solve in the same way as for the previous group of equations. Setting $\rho = a$ in Eq. (\ref{gij-Cyl2}) and then substituting into Eq. (\ref{g0j-Cyl2}) yields
\begin{align}
g ^{0} _{m m ^{\prime} , j} (\rho , \rho ^{\prime}) & = - i \tilde{\alpha} \left< v _{j} \right> _{m m ^{\prime}} \, \mathfrak{g} _{m} (\rho , a) g _{m ^{\prime}} (a , \rho ^{\prime}) \notag \\ & \phantom{=} - \tilde{\alpha} ^{2} \, \mathfrak{f} _{m} (k) \, \mathfrak{g} _{m} (\rho , a) \, g ^{0} _{m m ^{\prime} , j} (a , \rho ^{\prime}) , \label{g0j-Cyl4} 
\end{align}
where the function $\mathfrak{f} _{m} (k)$ is given by Eq. (\ref{f-function-Cyl}). Solving for $g ^{0} _{m m ^{\prime} , j} (a , \rho ^{\prime})$ by setting $\rho = a$ in Eq. (\ref{g0j-Cyl4}) and inserting the result back in this equation, we obtain
\begin{align}
g ^{0} _{m m ^{\prime} , j} (\rho , \rho ^{\prime}) &= \frac{- i \tilde{\alpha} \left< v _{j} \right> _{m m ^{\prime}}}{1 + \tilde{\alpha} ^{2} \mathfrak{f} _{m} (k)  \mathfrak{g} _{m} (a,a)}  \mathfrak{g} _{m} (\rho , a) \, g _{m ^{\prime}} (a , \rho ^{\prime}) . \label{g0j-CylFIN} 
\end{align}
The remaining components can be computed similarly. The substitution of $g ^{0} _{m m ^{\prime} , j} (a , \rho ^{\prime})$ in Eq. (\ref{gij-Cyl2}) yields
\begin{align}
g ^{i} _{m m ^{\prime} , j} (\rho , \rho ^{\prime}) &= \eta ^{i} _{\phantom{i}j} \delta _{mm ^{\prime}} g _{m} (\rho , \rho ^{\prime}) + \tilde{\alpha} ^{2} g _{m} (\rho , a) g _{m ^{\prime}} (a , \rho ^{\prime}) \notag \\ & \phantom{=} \times \sum _{m ^{\prime \prime}} \frac{\left< v ^{i} \right> _{m m ^{\prime \prime}} \left< v _{j} \right> _{m ^{\prime \prime} m ^{\prime}} \mathfrak{g} _{m ^{\prime \prime}} (a,a)}{1 + \tilde{\alpha} ^{2} \mathfrak{f} _{m ^{\prime \prime}} (k)  \mathfrak{g} _{m ^{\prime \prime}} (a,a)} . \label{gij-CylFIN} 
\end{align}
We observe that Eqs. (\ref{g00-CylFIN})-(\ref{gi0-CylFIN}) and Eqs. (\ref{g0j-CylFIN})-(\ref{gij-CylFIN}) contain all the required elements of the GF matrix. With the help of these, we can solve for the electromagnetic fields produced by arbitrary configuration sources in interaction with a real 3D cylindrical topological insulator.

Finally, the reciprocity between the position of the unit charge and the position at which the GF is evaluated, $G _{\mu \nu} (\textbf{r},\textbf{r} ^{\prime}) = G _{\nu \mu} (\textbf{r} ^{\prime} , \textbf{r})$, demands
\begin{align}
g _{mm ^{\prime}, \mu \nu} (\rho , \rho ^{\prime} ; k) = g _{m ^{\prime} m, \nu \mu} (\rho ^{\prime}, \rho ; - k) ,
\end{align}
which we verify directly from Eqs. (\ref{g00-CylFIN}), (\ref{gi0-CylFIN}), (\ref{g0j-CylFIN}) and (\ref{gij-CylFIN}).

\section{Applications} \label{Application}

\subsection{Line charge outside a cylindrical TI}

\begin{figure}[tbp]
\begin{center}
\includegraphics[scale=0.8]{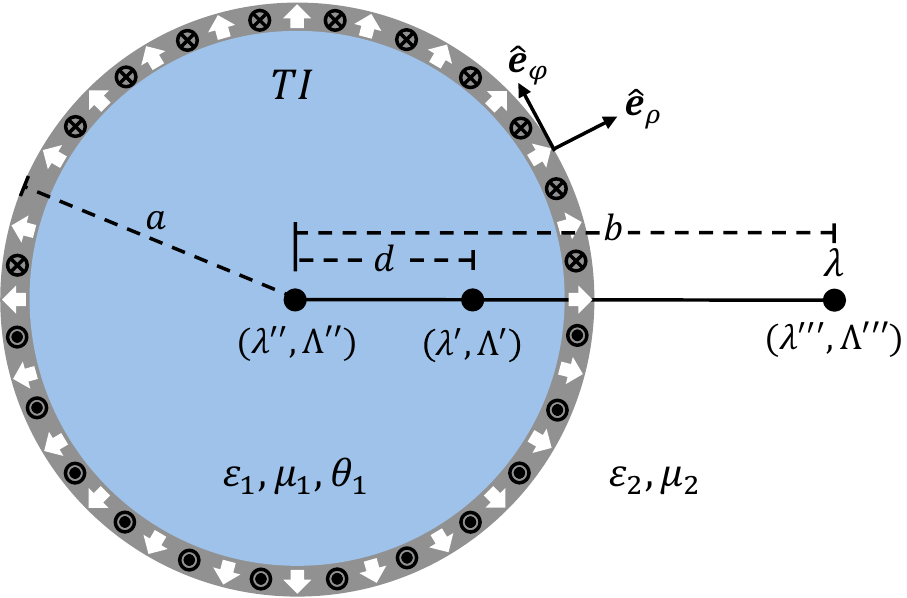}
\end{center}
\caption{{\protect\small Illustration of the images electric and magnetic charge densities induced by a line charge near a cylindrical TI. A magnetic layer is deposited on the surface, as indicated by the gray layer with white arrows. The symbols $\otimes$ and $\odot$ indicate the direction of the surface Hall current for $\lambda > 0$.}}
\label{Charged-Cyl-TI}
\end{figure}

We first consider the case of a cylindrical topological insulator of radius $a$, permittivity $\epsilon _{1}$ and axion angle $\theta _{1}$, when an infinite straight wire carrying charge density $\lambda$ is located in a dielectric fluid with permittivity $\epsilon _{2}$ at a distance $b > a$ from the center of the cylinder, as shown in Fig. \ref{Charged-Cyl-TI}. We choose the coordinates such that $\varphi ^{\prime} = 0$. Therefore, the current density is $J ^{\mu} (\textbf{r} ^{\prime}) = \frac{\lambda c}{b} \eta ^{\mu} _{\phantom{\mu}0} \delta (\varphi ^{\prime}) \delta (\rho ^{\prime} - b)$. The EM four-potential for this problem can thus be written as
\begin{align}
A ^{\mu} (\textbf{r}) = 2 \lambda \lim _{k \rightarrow 0} \sum _{m,m ^{\prime}} g ^{\mu} _{mm ^{\prime},0} (\rho , b ) \, e ^{im \varphi} , \label{SolLineChargeCyl}
\end{align}
where the various components of the reduced GF are given by Eqs. (\ref{g00-CylFIN}) and (\ref{gi0-CylFIN}), with $\tilde{\alpha} = - \alpha \theta _{1} / \pi$. Using the symmetry properties $\mathfrak{g} _{m} = \mathfrak{g} _{- m}$ and $g _{m} = g _{-m}$ of the reduced free GFs we find that the scalar and the (nonzero component of the) vector potential takes the form
\begin{align}
\phi (\textbf{r}) &=  2 \lambda \lim _{k \rightarrow 0} \left\lbrace \mathfrak{g} _{0} (\rho , b) + 2 \sum _{m = 1 } ^{+ \infty} \Bigg[ \mathfrak{g} _{m} (\rho , b) \right. \notag \\[2pt] & \phantom{=} \left.  - \frac{\tilde{\alpha} ^{2} \mathfrak{f} _{m} (k) \mathfrak{g} _{m} (\rho , a) \mathfrak{g} _{m} (a , b)}{1 + \tilde{\alpha} ^{2} \mathfrak{f} _{m} (k) \mathfrak{g} _{m} (a,a)} \Bigg] \cos \left( m \varphi \right) \right\rbrace , \notag \\[2pt]  A _{z} (\textbf{r}) &=  4 \lambda \tilde{\alpha} \lim _{k \rightarrow 0} \sum _{m = 1} ^{+ \infty} \frac{m g _{m} (\rho , a ) \mathfrak{g} _{m} ( a , b) \sin \left( m \varphi \right)}{1 + \tilde{\alpha} ^{2} \mathfrak{f} _{m} (k) \mathfrak{g} _{m} (a,a)} , \label{em-pots}
\end{align}
where we have also used that $\lim _{k \rightarrow 0} \mathfrak{f} _{0} (k) = 0$. The reduced free GFs, $\mathfrak{g} _{m}$ and $g _{m}$, for $\rho ^{\prime} > a$, are derived in the Appendix. Using the limiting form of the modified Bessel functions for small arguments \cite{Gradshteyn}, the limit $k \rightarrow 0$ of the reduced free GFs for $m>0$ becomes
\begin{align}
\lim _{k \rightarrow 0} \mathfrak{g} _{m} (\rho , \rho ^{\prime}) &= \frac{1}{2 m \epsilon _{2}} \left[ \left( \frac{\rho _{<}}{\rho _{>}} \right) ^{m} + \frac{\epsilon _{2} - \epsilon _{1}}{\epsilon _{2} + \epsilon _{1}} \left( \frac{a _{<}}{\rho _{>}} \frac{a}{\rho ^{\prime}} \right) ^{m} \right] , \notag \\ \lim _{k \rightarrow 0} g _{m} (\rho , \rho ^{\prime}) &= \frac{1}{2m} \left( \frac{\rho _{<}}{\rho _{>}} \right) ^{m} , \label{em-pots2}
\end{align}
and therefore $\lim _{k \rightarrow 0} \mathfrak{f} _{m} (k) = m /2$. Using the previous results the EM potentials become
\begin{align}
\phi (\textbf{r}) &= - \frac{2 \lambda}{\epsilon _{2}} \log \rho _{>}  + \frac{2 \lambda}{\epsilon _{2}} \sum _{m = 1 } ^{+ \infty} \Bigg[ \frac{1}{m} \left( \frac{b _{<}}{\rho _{>}} \right) ^{m} \notag \\[2pt] & \phantom{=} + \frac{2(\epsilon _{2} - \epsilon _{1}) - \tilde{\alpha} ^{2}}{2(\epsilon _{2} + \epsilon _{1}) + \tilde{\alpha} ^{2}} \frac{1}{m} \left( \frac{a _{<}}{\rho _{>}} \frac{a}{b} \right) ^{m} \Bigg] \, \cos \left( m \varphi \right) , \notag \\[2pt] A _{z} (\textbf{r}) &= \frac{4 \lambda \tilde{\alpha}}{2(\epsilon _{2} + \epsilon _{1}) + \tilde{\alpha} ^{2}} \sum _{m = 1} ^{+ \infty} \left( \frac{a _{<}}{\rho _{>}} \frac{a}{b} \right) ^{m} \frac{\sin \left( m \varphi \right) }{m},  \label{em-pots3}
\end{align}
where $x _{<} / y _{>}$ denotes the ratio between the lesser and the greater between $x$ and $y$. The summations can be performed analytically, and the results can be written in the simple form
\begin{align}
\phi (\textbf{r}) &= - \frac{\lambda}{\epsilon _{2}} \log \left( \rho ^{2} + b ^{2} - 2 \rho b \cos \varphi \right) \notag \\ & \phantom{=} - \frac{\lambda}{\epsilon _{2}} \frac{2(\epsilon _{2} - \epsilon _{1}) - \tilde{\alpha} ^{2}}{2(\epsilon _{2} + \epsilon _{1}) + \tilde{\alpha} ^{2}} \log \left( 1 +  r ^{2} - 2 r \cos \varphi \right) , \notag \\ A _{z} (\textbf{r}) &= \frac{4 \lambda \tilde{\alpha}}{2(\epsilon _{2} + \epsilon _{1}) + \tilde{\alpha} ^{2}} \arctan \left( \frac{ r \sin \varphi}{1 - r \cos \varphi} \right) , \label{EM-LineCyl}
\end{align}
with the notation $r \equiv ( a _{<} / \rho _{>} ) ( a / b )$.  
Next, we analyze the field strengths. The EM potentials in the dielectric fluid (outside the TI, $\rho > a$) are given by Eq. (\ref{EM-LineCyl}), with $r = d / \rho$ and $d = a ^{2} / b$. The corresponding electric and magnetic fields can be calculated directly, with the result
\begin{align}
\textbf{E} _{\mbox{\scriptsize{o}}} (\textbf{r}) & = \frac{2 \lambda}{\epsilon _{2}} \frac{( \rho - b \cos \varphi ) \hat{\textbf{e}} _{\rho} + b \sin \varphi \hat{\textbf{e}} _{\varphi}}{\rho ^{2} + b ^{2} - 2 \rho b \cos \varphi} + \frac{2  \lambda ^{\prime \prime}}{\epsilon _{2}} \frac{\hat{\textbf{e}} _{\rho}}{\rho} \notag \\[2pt] & \phantom{=} + \frac{2 \lambda ^{\prime}}{\epsilon _{2}} \frac{( \rho - d \cos \varphi ) \hat{\textbf{e}} _{\rho} + d \sin \varphi \hat{\textbf{e}} _{\varphi}}{\rho ^{2} + d ^{2} - 2 \rho d \cos \varphi} , \notag \\ \textbf{B} _{\mbox{\scriptsize{o}}} (\textbf{r}) &= 2 \Lambda ^{\prime} \frac{( \rho - d \cos \varphi ) \hat{\textbf{e}} _{\rho} + d \sin \varphi \hat{\textbf{e}} _{\varphi}}{\rho ^{2} + d ^{2} - 2 \rho d \cos \varphi} + 2 \Lambda ^{\prime \prime} \frac{\hat{\textbf{e}} _{\rho}}{\rho} , \label{B-Field-RI-Cyl}
\end{align}
where
\begin{align}
\lambda ^{\prime} = \lambda \frac{2(\epsilon _{2} - \epsilon _{1}) - \tilde{\alpha} ^{2}}{2(\epsilon _{2} + \epsilon _{1}) + \tilde{\alpha} ^{2}} , \quad \Lambda ^{\prime} = \frac{2 \lambda \tilde{\alpha}}{2(\epsilon _{2} + \epsilon _{1}) + \tilde{\alpha} ^{2}} . \label{densities}
\end{align}
These fields can be interpreted as follows. The electric field outside the TI corresponds to that generated by the original line charge, with electric charge density $\lambda$ and centered at $b$, plus two image parallel wires: one of strength $\lambda ^{\prime \prime} = - \lambda ^{\prime}$ at the origin and the other of strength $\lambda ^{\prime}$ located at $d = a ^{2} / b$. The magnetic field resembles that produced by two infinite straight wires which carry uniform magnetic charges per unit length: one of strength $\Lambda ^{\prime \prime} = - \Lambda ^{\prime}$ at the origin and the other of strength $\Lambda ^{\prime}$ centered at $d$. See Fig. \ref{Charged-Cyl-TI}.

Inside the TI ($\rho < a$) the EM potentials are given by Eq. (\ref{EM-LineCyl}) with $r = \rho / b$, wherefrom we derive the EM fields
\begin{align}
\textbf{E} _{\mbox{\scriptsize{i}}} (\textbf{r}) &= \frac{2 \lambda ^{\prime \prime \prime}}{\epsilon _{2}} \frac{( \rho - b \cos \varphi ) \hat{\textbf{e}} _{\rho} + b \sin \varphi \hat{\textbf{e}} _{\varphi}}{\rho ^{2} + b ^{2} - 2 \rho b \cos \varphi} , \notag \\ \textbf{B} _{\mbox{\scriptsize{i}}} (\textbf{r}) &= 2 \Lambda ^{\prime \prime \prime} \frac{( \rho - b \cos \varphi ) \hat{\textbf{e}} _{\rho} + b \sin \varphi \hat{\textbf{e}} _{\varphi}}{\rho ^{2} + b ^{2} - 2 \rho b \cos \varphi} ,
\end{align}
with $\lambda ^{\prime \prime \prime} = \lambda + \lambda ^{\prime}$ and $\Lambda ^{\prime \prime \prime} = \Lambda ^{\prime \prime}$. We observe that, inside the TI, the electric field corresponds to that produced by a uniformly charged wire of strength $\lambda ^{\prime \prime \prime}$ at $\rho = b$; while the magnetic field is equivalent to the one produced by an infinite straight wire carrying a uniform magnetic charge density $\Lambda ^{\prime \prime \prime}$, also at $\rho = b$. Note that in the limiting case $\theta _{1} = 0$, we obtain that $\textbf{B} _{\mbox{\scriptsize{o}}} = \textbf{B} _{\mbox{\scriptsize{i}}} = \textbf{0}$ due to the absence of the TME, and we recover correctly the electric field reported for a line charge outside a dielectric cylinder \cite{Smythe}.

The results of the present example are quite interesting. When a charged line is placed outside of a cylindrical topological insulator, of which the surface states have been gapped by TR symmetry breaking, the charge density polarizes and magnetizes the TI due to the TME effect. In this situation, the TI can be described by the appearance of infinite straight wires which carry uniform electric and magnetic charges inside the material, as shown in Fig. \ref{Charged-Cyl-TI}. This effect can be thought as an extension of the image magnetic monopole effect \cite{Qi-Monopole}. The wires, carrying a uniform charge density of magnetic monopoles, are unique signatures of the nontrivial topology of the bulk, and they are sourced by the surface Hall current density (in the linear response regime)
\begin{align}
\textbf{J} _{\mbox{\scriptsize Hall}} = - \frac{\eta \, \sigma _{\mbox{\scriptsize Hall}}}{\epsilon _{1} + \epsilon _{2}} \frac{4 \lambda \sin \varphi}{b + d - 2 a \cos \varphi} \; \hat{\textbf{e}} _{z} , \label{HallLineaCyl}
\end{align}
where $\sigma _{\mbox{\scriptsize Hall}} = (n + 1/2) \frac{e ^{2}}{h}$ is the quantized Hall conductivity and $\eta \equiv \mbox{sgn} \left( \textbf{M} \cdot \hat{\textbf{e}} _{\rho} \right) = \pm 1$. The direction of the surface magnetization, $\textbf{M}$, determines the sign of $\theta _{1}$: $+$ ($-$) if it points out (in) to the TI surface \cite{Qi-TFT}. For the configuration depicted in Fig. \ref{Charged-Cyl-TI} we have $\eta = + 1$. From Eq. (\ref{HallLineaCyl}) we observe that the surface Hall current splits into two longitudinal branches flowing in opposite directions, as dictated by $\sin \varphi$ in Eq. (\ref{HallLineaCyl}). A similar conclusion was reached in Ref. \cite{Campos}, where the authors study the geometric reversion of the Hall current in spherical and cylindrical cavities. In Fig. \ref{Charged-Cyl-TI} we indicate the direction of the surface Hall current for $\lambda > 0$ with the symbols $\odot$ (out of the page) and $\otimes$ (into the page). Noticeably, the sign of the surface current can be reversed by inverting the direction of the surface magnetization. In the last section \ref{Discussion} we discuss two posible experimental routes to detect the TME in this system, namely, the Lorentz force and the magnetic force miscrocopy.

\subsection{Line charge in a cylindrical cavity}
\begin{figure}[tbp]
\begin{center}
\includegraphics[scale=0.8]{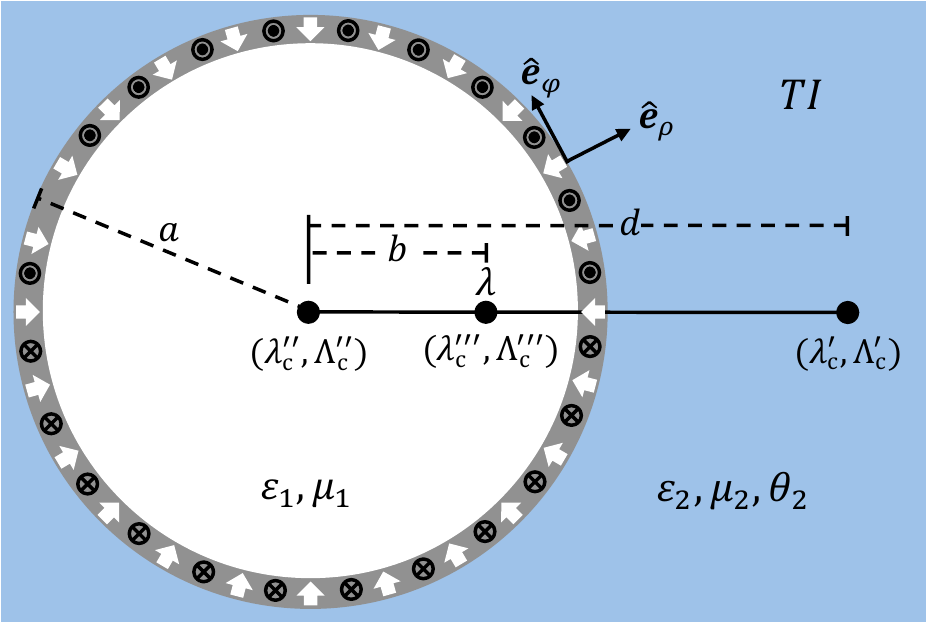}
\end{center}
\caption{{\protect\small Illustration of the images electric and magnetic charge densities induced by a line current placed in a cylindrical cavity. The TI surface is covered with a thin magnetic layer (gray layer) and the symbols $\otimes$ and $\odot$ indicate the direction of the surface Hall current for $\lambda > 0$.}}
\label{Charged-CylCav-TI}
\end{figure}
Now let us consider a TI (with permittivity $\epsilon _{2}$ and axion angle $\theta _{2}$) with a cylindrical dielectric-filled cavity (of radius $a$ and permittivity $\epsilon _{1}$), as shown in Fig. \ref{Charged-CylCav-TI}. A straight wire of charge density $\lambda$ is placed in the cavity parallel to the symmetry axis of the cylinder and at a distance $b < a$ from this. Now, $\tilde{\alpha} = \alpha \theta _{2} / \pi$. We can solve this problem in a similar way to the previous example. Again, we choose the coordinates in which $\varphi ^{\prime} = 0$, such that the current density is $J ^{\mu} (\textbf{r} ^{\prime}) = \frac{\lambda c}{b} \eta ^{\mu} _{\phantom{\mu}0} \delta (\varphi ^{\prime}) \delta (\rho ^{\prime} - b)$. The EM potentials are still given by Eq. (\ref{em-pots}) with, however, the reduced free GF $\mathfrak{g} _{m}$ given by Eqs. (\ref{gfrho'<a1}) and (\ref{gfrho'<a2}) for $\rho ^{\prime} < a$. One can further see that the limit $k \rightarrow 0$ of this reduced free GF for $m > 0$ is
\begin{align}
\lim _{k \rightarrow 0} \mathfrak{g} _{m} (\rho , \rho ^{\prime}) &= \frac{1}{2 m \epsilon _{1}} \left[ \left( \frac{\rho _{<}}{\rho _{>}} \right) ^{m} + \frac{\epsilon _{1} - \epsilon _{2}}{\epsilon _{1} + \epsilon _{2}} \left( \frac{\rho _{<}}{a _{>}} \frac{\rho ^{\prime}}{a} \right) ^{m} \right] ,
\end{align}
and therefore the EM potentials are likewise given by Eq. (\ref{EM-LineCyl}) albeit with the replacements $\epsilon _{1,2} \rightarrow \epsilon _{2,1}$ and $r \rightarrow t = ( \rho _{<} / a _{>} ) ( b / a )$. As before, the EM potentials are not illuminating, but the fields do. The electric and magnetic fields outside the TI (in the cavity, $\rho < a$) are
\begin{align}
\textbf{E} _{\mbox{\scriptsize{o}}} (\textbf{r}) &= \frac{2 \lambda}{\epsilon _{1}} \frac{( \rho - b \cos \varphi ) \hat{\textbf{e}} _{\rho} + b \sin \varphi \hat{\textbf{e}} _{\varphi}}{\rho ^{2} + b ^{2} - 2 \rho b \cos \varphi} + \notag \\[2pt] & \phantom{=} + \frac{2 \lambda ^{\prime} _{\mbox{\scriptsize c}}}{\epsilon _{1}} \frac{( \rho - d \cos \varphi ) \hat{\textbf{e}} _{\rho} + d \sin \varphi \hat{\textbf{e}} _{\varphi}}{\rho ^{2} + d ^{2} - 2 \rho d \cos \varphi} , \notag \\ \textbf{B} _{\mbox{\scriptsize{o}}} (\textbf{r}) &= 2 \Lambda ^{\prime} _{\mbox{\scriptsize c}} \frac{( \rho - d \cos \varphi ) \hat{\textbf{e}} _{\rho} + d \sin \varphi \hat{\textbf{e}} _{\varphi}}{\rho ^{2} + d ^{2} - 2 \rho d \cos \varphi} ,
\end{align}
where
\begin{align}
\lambda ^{\prime} _{\mbox{\scriptsize c}} = \lambda \frac{2(\epsilon _{1} - \epsilon _{2}) - \tilde{\alpha} ^{2}}{2(\epsilon _{1} + \epsilon _{2}) + \tilde{\alpha} ^{2}} , \quad \Lambda ^{\prime} _{\mbox{\scriptsize c}} = \Lambda ^{\prime \prime} .
\end{align}
As before, we interpret the EM fields in terms of images. The electric field corresponds to that generated by the original line charge, with electric charge density $\lambda$ at $\rho = b$, plus one image parallel wire of strength $\lambda ^{\prime} _{\mbox{\scriptsize c}}$ at $d = a ^{2} / b$. The magnetic field resembles to that produced by an image straight wire carrying magnetic charge density $\Lambda ^{\prime} _{\mbox{\scriptsize c}}$ at $\rho = d$.

The EM fields inside the TI ($\rho > a$) are found to be
\begin{align}
\textbf{E} _{\mbox{\scriptsize{i}}} (\textbf{r}) &= \frac{2 \lambda ^{\prime \prime \prime} _{\mbox{\scriptsize c}}}{\epsilon _{1}} \frac{( \rho - b \cos \varphi ) \hat{\textbf{e}} _{\rho} + b \sin \varphi \hat{\textbf{e}} _{\varphi}}{\rho ^{2} + b ^{2} - 2 \rho b \cos \varphi} + \frac{2 \lambda ^{\prime \prime} _{\mbox{\scriptsize c}}}{\epsilon _{1}} \frac{\hat{\textbf{e}} _{\rho} }{\rho} , \notag \\ \textbf{B} _{\mbox{\scriptsize{i}}} (\textbf{r}) &= 2 \Lambda ^{\prime \prime \prime} _{\mbox{\scriptsize c}} \frac{( \rho - b \cos \varphi ) \hat{\textbf{e}} _{\rho} + b \sin \varphi \hat{\textbf{e}} _{\varphi}}{\rho ^{2} + b ^{2} - 2 \rho b \cos \varphi} + 2 \Lambda ^{\prime \prime} _{\mbox{\scriptsize c}} \frac{\hat{\textbf{e}} _{\rho}}{\rho} , \label{Fields-CylCav-Inside}
\end{align}
where $\Lambda ^{\prime \prime \prime} _{\mbox{\scriptsize c}} = - \Lambda ^{\prime} _{\mbox{\scriptsize c}}$ and $\Lambda ^{\prime \prime} _{\mbox{\scriptsize c}} = \Lambda ^{\prime} _{\mbox{\scriptsize c}}$. The electric field inside the TI is as if produced by two infinitely uniformly charged wires, one of strength $\lambda ^{\prime \prime \prime} _{\mbox{\scriptsize c}}$ at $\rho = b$, and the other of strength $\lambda ^{\prime \prime} _{\mbox{\scriptsize c}}$ at the origin. The magnetic field is due to two image straight wires carrying magnetic charge densities: one of strength $\Lambda ^{\prime \prime} _{\mbox{\scriptsize c}}$ at the origin and the other of strength $\Lambda ^{\prime \prime \prime} _{\mbox{\scriptsize c}}$ at $\rho = b$. See Fig. \ref{Charged-CylCav-TI}.

A simple calculation show us that the surface Hall current which sources the monopole-like magnetic fields is also given by Eq. (\ref{HallLineaCyl}). However, the difference is the sign of the surface magnetization: in the previous example $\textbf{M}$ points outward the symmetry axis, while in the present case $\textbf{M}$ is pointing inward. This means that, as in the previous example, the surface Hall current splits into two longitudinal branches flowing in opposite directions. However, in this case, the surface Hall current flows in the positive (negative) $z$-direction for $0 < \varphi < \pi$ ($\pi < \varphi < 2 \pi$), as shown in Fig. \ref{Charged-CylCav-TI}.

\subsection{Line current outside a cylindrical TI}
\begin{figure}[tbp]
\begin{center}
\includegraphics[scale=0.8]{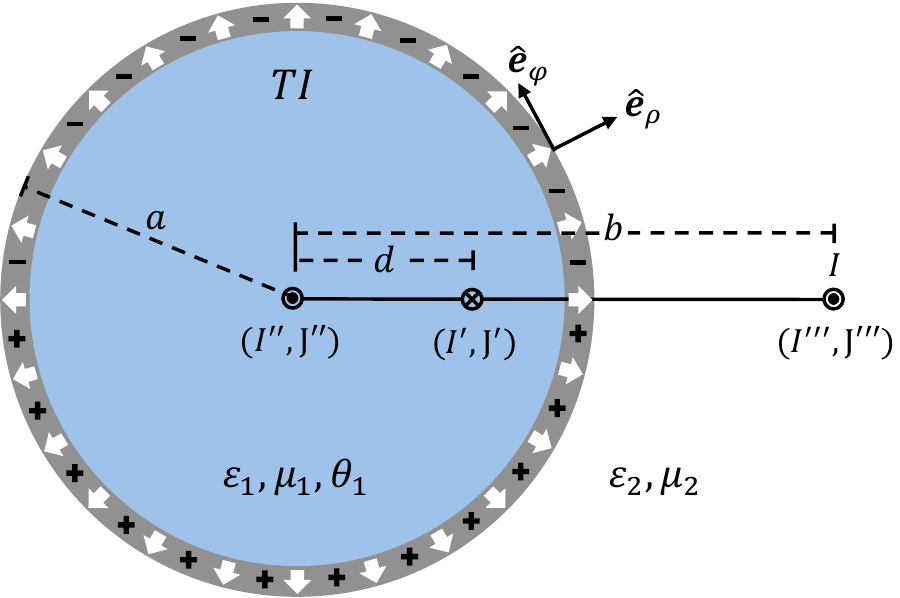}
\end{center}
\caption{{\protect\small Illustration of the images electric and magnetic current densities induced by a line current near a cylindrical TI. The plus and minus signs in the surface layer indicate the sign of the anomalous charge density sourcing the magnetic currents for $I > 0$.}}
\label{Current-Cyl-TI}
\end{figure}
An obvious extension of the previous examples consists in replacing the line charges by line currents either in front of a cylindrical TI or in a cylindrical cavity. Let us first consider an infinite straight current-carrying wire parallel to the $z$-axis carrying a current $I$ in the $+z$ direction. The wire is located in a dielectric fluid (with permittivity $\epsilon _{2}$) in front of a cylindrical TI (of radius $a$, permittivity $\epsilon _{1}$ and axion angle $\theta _{1}$) at a distance $b > a$ from the $z$-axis, as shown in Fig. \ref{Current-Cyl-TI}. Choosing the coordinates such that $\varphi ^{\prime} = 0$, the current density becomes $J ^{\mu} (\textbf{r} ^{\prime}) = \frac{I}{b} \eta ^{\mu} _{\phantom{\mu}3} \delta (\varphi ^{\prime}) \delta (\rho ^{\prime} - b)$, and Eq. (\ref{EMPotSol}) can be integrated to obtain
\begin{align}
A ^{\mu} (\textbf{r}) = \frac{2I}{c} \lim _{k \rightarrow 0} \sum _{m,m ^{\prime}} g ^{\mu} _{mm ^{\prime},3} (\rho , b ) \, e ^{im \varphi} , \label{SolLineCurrentCyl}
\end{align}
where the required components of the reduced GF are given by Eqs. (\ref{g0j-CylFIN}) and (\ref{gij-CylFIN}). With the use of these components, the scalar and vector potentials become
\begin{align}
\phi (\textbf{r}) &= - \frac{4 \tilde{\alpha} I}{c} \lim _{k \rightarrow 0} \sum _{m = 1} ^{+ \infty} \frac{m \mathfrak{g} _{m} (\rho , a ) g _{m} ( a , b) \sin \left( m \varphi \right)}{1 + \tilde{\alpha} ^{2} \mathfrak{f} _{m} (k) \mathfrak{g} _{m} (a,a)} , \notag \\[2pt]  A _{z} (\textbf{r}) &= \frac{2 I}{c} \lim _{k \rightarrow 0} \left\lbrace g _{0} (\rho , b) + 2 \sum _{m = 1 } ^{+ \infty} \Bigg[ g _{m} (\rho , b) \right. \notag \\[2pt] & \phantom{=} \left.  - \frac{\tilde{\alpha} ^{2} m ^{2} \mathfrak{g} _{m} (a,a) g _{m} (\rho , a) g _{m} (a , b)}{1 + \tilde{\alpha} ^{2} \mathfrak{f} _{m} (k) \mathfrak{g} _{m} (a,a)} \Bigg] \cos \left( m \varphi \right) \right\rbrace . \label{em-potscurrent}
\end{align}
The explicit form of these potentials can be calculated in the same way as in the previous examples: we first substitute the limiting form of the reduced free GFs, $\mathfrak{g} _{m} (\rho , \rho ^{\prime})$ $g _{m} (\rho , \rho ^{\prime})$, given by Eq. (\ref{em-pots2}), into Eq. (\ref{em-potscurrent}), and then we compute the resulting summations. The final result is
\begin{align}
\phi (\textbf{r}) &= - \frac{4 I \tilde{\alpha} / c}{2(\epsilon _{2} + \epsilon _{1}) + \tilde{\alpha} ^{2}} \arctan \left( \frac{ r \sin \varphi}{1 - r \cos \varphi} \right) , \notag \\ A _{z} (\textbf{r}) &= - \frac{I}{c} \log \left( \rho ^{2} + b ^{2} - 2 \rho b \cos \varphi \right) \notag \\ & \phantom{=} + \frac{\tilde{\alpha} ^{2} I / c }{2(\epsilon _{2} + \epsilon _{1}) + \tilde{\alpha} ^{2}} \log \left( 1 +  r ^{2} - 2 r \cos \varphi \right) , \label{EM-CurrentLineCyl}
\end{align}
where $r = ( a _{<} / \rho _{>} ) (a / b)$. Now we analyze the field strengths, which derives from these potentials in the usual way. Outside the TI ($\rho > a$), the EM fields become
\begin{align}
\textbf{E} _{\mbox{\scriptsize{o}}} (\textbf{r}) & = \frac{2 J ^{\prime}}{c} \frac{ (\rho - d \cos \varphi) \hat{\textbf{e}} _{\varphi} - d \sin \varphi \hat{\textbf{e}} _{\rho}}{\rho ^{2} + d ^{2} - 2 \rho d \cos \varphi} + \frac{2 J ^{\prime \prime}}{c} \frac{\hat{\textbf{e}} _{\varphi}}{\rho} , \notag \\ \textbf{B} _{\mbox{\scriptsize{o}}} (\textbf{r}) &= \frac{2I}{c} \frac{(\rho - b \cos \varphi) \hat{\textbf{e}} _{\varphi} - b \sin \varphi \hat{\textbf{e}} _{\rho}}{\rho ^{2} + b ^{2} - 2 \rho b \cos \varphi} + \frac{2 I ^{\prime \prime}}{c} \frac{\hat{\textbf{e}} _{\varphi}}{\rho} \notag \\ & \phantom{=} + \frac{2I ^{\prime}}{c} \frac{(\rho - d \cos \varphi) \hat{\textbf{e}} _{\varphi} - d \sin \varphi \hat{\textbf{e}} _{\rho}}{\rho ^{2} + d ^{2} - 2 \rho d \cos \varphi}  , \label{B-Field-RI-CylCurrent}
\end{align}
where
\begin{align}
J ^{\prime} = \frac{2 \tilde{\alpha} I}{2 (\epsilon _{1} + \epsilon _{2}) + \tilde{\alpha} ^{2}} , \quad I ^{\prime \prime} = \frac{ \tilde{\alpha} ^{2} I}{2 (\epsilon _{1} + \epsilon _{2}) + \tilde{\alpha} ^{2}} , 
\end{align}
$J ^{\prime \prime} = - J ^{\prime}$ and $I ^{\prime} = - I ^{\prime \prime}$. We observe that the magnetic field corresponds to that generated by the original wire with current $I$ (flowing in the $+z$-direction) at $\rho = b $ plus two image electric currents: one of strength $I ^{\prime \prime}$ flowing in the same $+z$ direction and centered at the origin, and the other of strength $I ^{\prime}$ flowing in the opposite $-z$ direction and centered at $d = a ^{2} / b $. The electric field can be interpreted in terms of two image wires carrying magnetic currents: one of strength $J ^{\prime \prime}$ centered at the origin and the other of strength $J ^{\prime}$ centered at $d$. Interestingly, the flow direction of these magnetic currents depends on the direction of the surface magnetization as follows. Since $\tilde{\alpha} = - \alpha \theta _{1} / \pi$ and $\theta _{1} = \eta (2n+1) \pi$, with $\eta = \mbox{sgn} (\textbf{M} \cdot \hat{\textbf{e}} _{\rho}) = \pm 1$, therefore $J ^{\prime} \propto - I \eta$. This means that, for the configuration depicted in Fig. \ref{Current-Cyl-TI}, where the surface magnetization is pointing outward the TI and the original electric current flows in the $+z$ direction, the magnetic current $J ^{\prime \prime}$ at the origin flows in the same $+z$ direction, while the other, $J ^{\prime}$ at $\rho = d$, flows in the opposite $-z$ direction. These results could be relevant when designing an experiment to observe these astonishing currents of magnetic monopoles, since the direction of the magnetic currents can be tuned by means of the surface magnetization, and strikingly, the image currents are exactly the same when inverting both, the direction of the original current and the direction of $\textbf{M}$.

Inside the TI ($\rho < a$), the electric and magnetic fields are given by
\begin{align}
\textbf{E} _{\mbox{\scriptsize{i}}} (\textbf{r}) & = \frac{2 J ^{\prime \prime \prime} }{c}  \frac{ (\rho - b \cos \varphi) \hat{\textbf{e}} _{\varphi} - b \sin \varphi \hat{\textbf{e}} _{\rho}}{\rho ^{2} + b ^{2} - 2 \rho b \cos \varphi} , \notag \\ \textbf{B} _{\mbox{\scriptsize{i}}} (\textbf{r}) &= \frac{2I ^{\prime \prime \prime}}{c} \frac{(\rho - b \cos \varphi) \hat{\textbf{e}} _{\varphi} - b \sin \varphi \hat{\textbf{e}} _{\rho}}{\rho ^{2} + b ^{2} - 2 \rho b \cos \varphi}  , \label{B-Field-RI-CylCurrent2}
\end{align}
where $J ^{\prime \prime \prime} = - J ^{\prime}$ and $I ^{\prime \prime \prime} = I - I ^{\prime \prime}$. We observe that the magnetic field is due to a straight wire carrying an electric current $I ^{\prime \prime \prime}$ flowing in the $+z$ direction and centered at $\rho = b$. The electric field corresponds to that generated by a wire carrying a magnetic current of strenght $J ^{\prime \prime \prime}$ centered at $\rho = b$. As before, the direction of the magnetic current can be tuned by means of the surface magnetization. For the configuration depicted in Fig. \ref{Current-Cyl-TI}, such current is flowing in the $+z$ direction.

The appearance of magnetic currents seems to violate Faraday's law, which remained unaltered in the presence of the $\theta$ term. The physical origin of the electric fields is, however, the anomalous surface charge density
\begin{align}
\rho _{\mbox{\scriptsize anom}} = - \frac{\eta \, \sigma _{\mbox{\scriptsize Hall}}}{c ^{2}} \frac{2I \, \sin \varphi}{b + d - 2 a \cos \varphi} . \label{AnomalousCharge}
\end{align} 
Clearly, the sign of this charge density depends on the direction of the surface magnetization. For the configuration depicted in Fig. \ref{Current-Cyl-TI}, $\rho _{\mbox{\scriptsize anom}}$ is negative (positive) for $0 < \varphi < \pi$ ($\pi < \varphi < 2 \pi$). This fact is represented in the figure with the plus and minus signs in the surface layer. Noticeably, this sign is tunable by means of the surface magnetization in a similar fashion to the direction of the previously discussed magnetic currents. In the last section \ref{Discussion} we discuss an experimental setup to test TME in this system: angle-resolved measurement.

\subsection{Line current in a cylindrical cavity}

Now let us briefly discuss the case in which the line current is located in the cylindrical cavity. To obtain the EM fields we follow a similar procedure to that used in the previous examples. The resulting EM potentials in this case are given by Eq. (\ref{EM-CurrentLineCyl}) albeit with the replacements $\epsilon _{1,2} \rightarrow \epsilon _{2,1}$ and $r \rightarrow t = ( \rho _{<} / a _{>} ) ( b / a )$. The corresponding EM fields are obtained in the usual way. Outside the TI (in the cavity, $\rho < a$) we find
\begin{align}
\textbf{E} _{\mbox{\scriptsize{o}}} (\textbf{r}) & = \frac{2 J ^{\prime} _{\mbox{\scriptsize c}}}{c} \frac{(\rho - d \cos \varphi) \hat{\textbf{e}} _{\varphi} - d \sin \varphi \hat{\textbf{e}} _{\rho}}{\rho ^{2} + d ^{2} - 2 \rho d \cos \varphi} , \notag \\ \textbf{B} _{\mbox{\scriptsize{o}}} (\textbf{r}) &= \frac{2I}{c} \frac{(\rho - b \cos \varphi) \hat{\textbf{e}} _{\varphi} - b \sin \varphi \hat{\textbf{e}} _{\rho}}{\rho ^{2} + b ^{2} - 2 \rho b \cos \varphi} \notag \\ & \phantom{=} + \frac{2 I ^{\prime} _{\mbox{\scriptsize c}}}{c} \frac{(\rho - d \cos \varphi) \hat{\textbf{e}} _{\varphi} - d \sin \varphi \hat{\textbf{e}} _{\rho}}{\rho ^{2} + d ^{2} - 2 \rho d \cos \varphi} ,
\end{align}
and, inside the TI ($\rho > a$), we obtain
\begin{align}
\textbf{E} _{\mbox{\scriptsize{i}}} (\textbf{r}) & = \frac{2 J ^{\prime \prime \prime} _{\mbox{\scriptsize c}}}{c} \frac{ (\rho - b \cos \varphi) \hat{\textbf{e}} _{\varphi} - b \sin \varphi \hat{\textbf{e}} _{\rho}}{\rho ^{2} + b ^{2} - 2 \rho b \cos \varphi} + \frac{2 J ^{\prime \prime} _{\mbox{\scriptsize c}}}{c} \frac{\hat{\textbf{e}} _{\varphi}}{\rho} , \notag \\ \textbf{B} _{\mbox{\scriptsize{i}}} (\textbf{r}) &= \frac{2 I ^{\prime \prime \prime} _{\mbox{\scriptsize c}}}{c} \frac{(\rho - b \cos \varphi) \hat{\textbf{e}} _{\varphi} - b \sin \varphi \hat{\textbf{e}} _{\rho}}{\rho ^{2} + b ^{2} - 2 \rho b \cos \varphi} + \frac{2 I ^{\prime \prime} _{\mbox{\scriptsize c}}}{c} \frac{\hat{\textbf{e}} _{\varphi}}{\rho} ,
\end{align}
where $J ^{\prime} _{\mbox{\scriptsize c}} = - J ^{\prime}$, $I ^{\prime} _{\mbox{\scriptsize c}} = - I ^{\prime}$, $J ^{\prime \prime} _{\mbox{\scriptsize c}} = J ^{\prime} _{\mbox{\scriptsize c}}$, $J ^{\prime \prime \prime} _{\mbox{\scriptsize c}} = - J ^{\prime} _{\mbox{\scriptsize c}}$, $I ^{\prime \prime} _{\mbox{\scriptsize c}} = - I ^{\prime} _{\mbox{\scriptsize c}}$ and $I ^{\prime \prime \prime} _{\mbox{\scriptsize c}} = I + I ^{\prime} _{\mbox{\scriptsize c}}$. As before, these fields admits a simple interpretation in terms of images electric and magnetic current densities. The magnetic field in the cavity is produced by the original wire, carrying a current $I$ in the $+z$ direction and centered at $b$, plus an image wire of strength $I ^{\prime} _{\mbox{\scriptsize c}}$ centered at $d = a ^{2} / b$. The electric field is as produced by a wire carrying a magnetic current of strength $J ^{\prime} _{\mbox{\scriptsize c}}$ centered at $d$. Inside the TI, the electric field can be interpreted in terms of two image wires carrying magnetic currents: one of strength $J ^{\prime \prime \prime} _{\mbox{\scriptsize c}}$ centered at $b$ and the other of strength $J ^{\prime \prime} _{\mbox{\scriptsize c}}$ centered at the origin. The magnetic fields is as due to two image electric currents: one of strength $I ^{\prime \prime \prime} _{\mbox{\scriptsize c}}$ centered at $b$ and the other of strength $I ^{\prime \prime} _{\mbox{\scriptsize c}}$ centered at the origin. The anomalous surface charge density which sources the electric fields is given by Eq. (\ref{AnomalousCharge}), with however, the direction of the magnetization pointing inward the symmetry axis.

\section{Discussion} \label{Discussion}

The non existence of magnetic monopoles has intrigued physicists for decades. The elusiveness of this particle in classical electrodynamics is apparent because of the lack of electric-magnetic duality in Maxwell's equations. A persuasive argument on the existence of magnetic monopoles was first put forward by Dirac \cite{Dirac}, who showed that if they exist then the electric charge must be quantized. Many years later, Polyakov \cite{Polyakov} and 't Hooft \cite{tHooft} showed that any unified theory of particle physics necessarily containts magnetic monopoles.

While the existence of true magnetic monopoles has not yet been verified, the quest for these became a burning topic in condensed matter physics lately, both theoretically and experimentally, since a number of systems have been shown to provide intriguing analogues. An exciting breakthrough was the suggestion that in magnetic materials called spin-ice \cite{Bramwell} the elementary excitations have a magnetic charge and behave as magnetic monopoles. The recent development of topological phases of matter have provided a new arena in which magnetic monopoles also appear, for example, when an electric charge is brought near to the surface of a planar TI \cite{Qi-Monopole}. It should be stressed that such an image magnetic monopole is not a real elementary particle, but it comes to be an artificial excitation describing the physical effects associated to the half-integer quantum Hall effect taking place at the surface of the TI.

The authors in Ref. \cite{Qi-Monopole} proposed an experimental setup to measure the image magnetic monopole field by means of local probes sensitive to small magnetic fields. Nevertheless, this striking effect is quite difficult to be observed experimentally because of a number of screening effects which arises in realistic situations \cite{Zang}. A way out of these embarrassing effects is, on the one hand, to replace the point charge by a charged wire in parallel with the TI surface, and on the other hand, by using cylindrical TIs, which are much more feasible for experiments than planar TIs. This is why, in this paper, we investigate the topological magnetoelectric effect in cylindrical TIs and we solve for the electromagnetic fields in the presence of charged and current-carrying wires in parallel with the symmetry axis of the TI.

For the sake of generality, we construct the Green's function describing the EM response of two topological media separated by a cylindrical interface. This GF describes the magnetoelectric effect of both i) a cylindrical TI surrounded by a dielectric fluid and ii) a TI with a cylindrical dielectric-filled cavity. As a first application, we tackled the problem of a line charge deposited in a dielectric fluid, outside of a cylindrical TI, as depicted in Fig. \ref{Charged-Cyl-TI}. We find that the EM fields can be interpreted as follows. The electric field outside the TI is generated by the original line charge plus two image parallel wires located inside the material; while the magnetic field resembles that produced by two straight wires carrying magnetic charges per unit length located also inside the TI. Analogously, the electric (magnetic) field inside the TI can be interpreted in terms of a straight wire carrying electric (magnetic) charge density located at the position of the original source. See Fig. \ref{Charged-Cyl-TI} for further details. The wires, carrying a uniform charge density of magnetic monopoles, are unique signatures of the nontrivial topology of the bulk, and they are sourced by the surface Hall current density (\ref{HallLineaCyl}). We recall that for this effect to be observed, a TR-symmetry breaking is needed, which is achieved by coating the TI surface with a thin ferromagnetic layer. Let us briefly discuss two possible routes to detect the TME effect in this system: the Lorentz force and the magnetic force microscopy (MFM). 

\emph{Lorentz force.} The force (per unit length) on the line charge at $\rho = b$ in the presence of the cylindrical TI of radius $a < b$ can be computed in a simple fashion as that due to the electric field of the image wires with charge densities $\lambda ^{\prime \prime}$ and $\lambda ^{\prime}$ at the origin and at $d = a ^{2} / b$, respectively. The result is
\begin{align}
\textbf{F} = \frac{2 \lambda ^{2}}{\epsilon _{2}} \frac{2 (\epsilon _{2} - \epsilon _{1}) - \tilde{\alpha} ^{2}}{2 (\epsilon _{2} + \epsilon _{1}) + \tilde{\alpha} ^{2}} \frac{a ^{2}}{b( b ^{2} - a ^{2})} \hat{\textbf{e}} _{\rho} , \label{force}
\end{align}
wherefrom we observe that the force upon the line charge is dominated by the nontopological optical contribution provided $\epsilon _{1} \neq \epsilon _{2}$. In order to isolate the topological contribution one can embed the probe in a dielectric fluid with the same dielectric constant than that of the topological insulator \cite{PRA-Hydrogenlike}, i.e. $\epsilon _{1} = \epsilon _{2}$. In this case, the force (\ref{force}) is $\alpha ^{2} \approx 10 ^{-5}$ smaller than the force that a conducting cylinder exherts upon an identical line charge in vacuum. Although this result is in the verge of the current experimental accessibility, its detection is challenging because of the practical difficulty in achieving exactly $\epsilon _{1} = \epsilon _{2}$, as well as a full insulating bulk behavior in TIs.

\emph{Magnetic Force Microscopy.} When the TI surface is coated with a thin ferromagnetic layer (whose magnetization points normal to its surface) the distribution of the magnetic field outside of the TI is that given by the image wires carrying magnetic charge density (provided the Fermi level lies within the magnetic gap). This magnetic field will in turn modify the film magnetization whose profile now dependens on the topological and optical parameters, as well as the azimuthal angle $\varphi$. Therefore, magnetic force microscopy may be used to probe such a magnetization pattern \cite{Campos}.

As a second example, we considered the case of a line charge deposited in a dielectric-filled cylindrical cavity inside of a semi-infinite topological insulator, as shown in Fig. \ref{Charged-CylCav-TI}. Similar to the previous case, the EM fields can also be interpreted in terms of suitable image electric and magnetic current densities. Once again, the Lorentz force and the MFM could be useful to detect the magnetoelectric effect in this configuration.

We also considered the problem of a line current deposited in a dielectric fluid near to the surface of a cylindrical TI. Compared with the magnetic fields induced by a single monopole or by a line charge, many moving electrons could provide novel TME effects, facilitating the experimental observations. Similar interpretations can be made in this case. The magnetic field outside the TI is produced by the original line current plus two image current lines located inside the TI; while the electric field is produced by two image wires carrying magnetic currents. Inside the TI, the magnetic (electric) field is that produced by straight wire carrying electric (magnetic) charge density
located at the position of the original source. These currents of magnetic monopoles are sourced by the anomalous surface charge density (\ref{AnomalousCharge}). Next we describe a possible experimental setup in which these results could be tested.

\emph{Angle-resolved measurement} The Lorentz force acting upon the current wire has two contributions. On the one hand, the original current generates two image currents, which would support a Lorentz force acting on the original current itself. This electric contribution is given by
\begin{align}
\textbf{F} _{\mbox{\scriptsize e}} = \frac{2 I ^{2}}{c ^{2}} \frac{\tilde{\alpha} ^{2}}{2 (\epsilon _{2} + \epsilon _{1}) + \tilde{\alpha} ^{2}} \frac{a ^{2}}{b( b ^{2} - a ^{2})} \hat{\textbf{e}} _{\rho} , \label{force2}
\end{align}
which is pointing radially. On the other hand, the electric fields produced by the two image wires carrying magnetic currents also exert a force upon the original wire. This magnetic contribution is given by
\begin{align}
\textbf{F} _{\mbox{\scriptsize m}} = \frac{2 I ^{2}}{v c} \frac{2 \tilde{\alpha}}{2 (\epsilon _{2} + \epsilon _{1}) + \tilde{\alpha} ^{2}} \frac{a ^{2}}{b( b ^{2} - a ^{2})} \hat{\textbf{e}} _{\varphi} , \label{force3}
\end{align}
where $v$ is the velocity of the electrons in the wire. Note that this anomalous force is orthogonal to the electric contribution (\ref{force2}). As a result, these two effects can be distinguished easily. Note that $\vert \textbf{F} _{\mbox{\scriptsize m}} \vert / \vert \textbf{F} _{\mbox{\scriptsize e}} \vert = 2c / v \alpha \gg 1$, thus implying that the anomalous force (\ref{force3}) is considerably larger than the usual electric force (\ref{force2}). Experimentally, the required current can be provided by the steady electron beam emmiting from low-energy electron gun (low-energy electron diffraction, for instance). While drifting above the cylincrical TI surface, the anomalous force (\ref{force3}) would deflect tangentially the trajectory of the electron beam. In principle, this deflection should be traced by angle-resolved measurement.

As a last application we also considered the problem of a line current deposited in a dielectric-filled cylindrical cavity inside of a semi-infinite TI. Once again, the EM fields also admits a simple interpretation in term of image current densities of electric and magnetic charges. The anomalous force acting on the line current could also be detected by angle-resolved measurement. In this case, we can make electrons go through a finite (along the $z$-axis) cavity and perform the measurement of the corresponding deflection at the end of the cavity.

\acknowledgements

Useful discussions with L. F. Urrutia, M. Cambiaso and A. Cortijo are warmly appreciated. The author was supported by the CONACyT postdoctoral Grant No. 234774.

\appendix

\section{Derivation of the reduced GF} \label{AppCyl}

In this section we solve the differential equation (\ref{DielectricCylRedGreen}) for the reduced free GF $\mathfrak{g} _{m} (\rho , \rho ^{\prime})$, with the permittivity function $\epsilon (\rho) = \epsilon _{1} H (a - \rho) + \epsilon _{2} H (\rho - a)$, where $H (x)$ is the Heaviside function. In the following we will first solve the problem by assuming that $\rho ^{\prime} > a$ (that is, the source lies in the medium 2). The solutions to Eq. (\ref{DielectricCylRedGreen}), subject to the usual boundary conditions 
\begin{align}
\epsilon _{1} \frac{\partial \mathfrak{g} _{m}}{\partial \rho} \Big| _{\rho = a - 0} = \epsilon _{2} \frac{\partial \mathfrak{g} _{m}}{\partial \rho} \Big| _{\rho = a + 0} , \label{BCs-RedGreenCyl} \\ \frac{\partial \mathfrak{g} _{m}}{\partial \rho} \Big| _{\rho = \rho ^{\prime} + 0} - \frac{\partial \mathfrak{g} _{m}}{\partial \rho} \Big| _{\rho = \rho ^{\prime} - 0} =  - \frac{1}{\rho ^{\prime} \epsilon (\rho ^{\prime})} , \label{BCs2-RedGreenCyl}
\end{align}
and the continuity of $\mathfrak{g} _{m}$ at $\rho = a$ and $\rho = \rho ^{\prime}$, can be expressed in terms of the solutions, $\mbox{I} _{m} (k \rho)$ and $\mbox{K} _{m} (k \rho)$, of the homogeneous equation. Here, $\mbox{I} _{m} (k \rho)$ and $\mbox{K} _{m} (k \rho)$ are the modified Bessel functions of the first and second kind, respectively. The forms of the solutions in the three regions are as follows:
\begin{align}
\rho < a &: \;\; \mathfrak{g} _{m} (\rho , \rho ^{\prime}) = A (\rho ^{\prime}) \, \mbox{I} _{m} (k \rho) , \notag \\[2pt] a < \rho < \rho ^{\prime} &: \;\; \mathfrak{g} _{m} (\rho , \rho ^{\prime}) = B (\rho ^{\prime}) \, \mbox{I} _{m} (k \rho) + C (\rho ^{\prime}) \, \mbox{K} _{m} (k \rho) , \notag \\[2pt] \rho > \rho ^{\prime} &: \;\; \mathfrak{g} _{m} (\rho , \rho ^{\prime}) = D (\rho ^{\prime}) \, \mbox{K} _{m} (k \rho) , \label{RedGreen3Cyl}
\end{align}
where the single function $\mbox{I} _{m} (k \rho)$ for the region $\rho < a$ is required by the boundary condition that $\mathfrak{g} _{m}$ remain finite for $\rho \rightarrow 0$. Similarly, the single function $\mbox{K} _{m} (k \rho)$ for $\rho > \rho ^{\prime}$ guarantees the correct convergence of the reduced free GF for $\rho \rightarrow \infty$. Imposing the boundary conditions we obtain the system of equations 
\begin{align}
A \, \mbox{I} _{m} (k a) = B \, \mbox{I} _{m} (k a) + C \, \mbox{K} _{m} (k a) , \notag \\[2pt] \epsilon _{1} A \, \mbox{I} _{m} ^{\prime} (k a) = \epsilon _{2} \left[ B \, \mbox{I} _{m} ^{\prime} (k a) + C \, \mbox{K} _{m} ^{\prime} (k a) \right] , \notag \\[2pt] B \, \mbox{I} _{m} (k \rho ^{\prime}) + C \, \mbox{K} _{m} (k \rho ^{\prime}) = D \, \mbox{K} _{m} (k \rho ^{\prime}) , \notag \\[2pt]  B \, \mbox{I} _{m} ^{\prime} (k \rho ^{\prime}) + C \, \mbox{K} _{m} ^{\prime} (k \rho ^{\prime}) - D \, \mbox{K} _{m} ^{\prime} (k \rho ^{\prime}) = 1 /  (k \rho ^{\prime} \epsilon _{2}) ,
\end{align}
where the prime denotes derivative with respect to the argument. Solving this system of equations and inserting the result back into Eq. (\ref{RedGreen3Cyl}) we find for the reduced free GF, in the medium 1 ($\rho < a$),
\begin{align}
\mathfrak{g} _{m} (\rho , \rho ^{\prime}) = \frac{1}{ka} \frac{\mbox{I} _{m} (k \rho) \, \mbox{K} _{m} (k \rho ^{\prime})}{W _{m} (k a)} , \label{gfrho'<a}
\end{align}
and, in the medium 2 ($\rho > a$), 
\begin{align}
\mathfrak{g} _{m} (\rho , \rho ^{\prime}) &= \frac{1}{\epsilon _{2}} \mbox{I} _{m} (k \rho _{<}) \, \mbox{K} _{m} (k \rho _{>}) + \mbox{K} _{m} (k \rho) \, \mbox{K} _{m} (k \rho ^{\prime}) \notag \\ & \phantom{=} \times  \frac{\mbox{I} _{m} (k a)}{\mbox{K} _{m} (k a)} \left[ \frac{1}{ka} \frac{1}{W _{m} (k a)} - \frac{1}{\epsilon _{2}} \right] , \label{gfrho'<a}
\end{align}
where $\rho _{>}$ ($\rho _{<}$) is the greater (lesser) of $\rho$ and $\rho ^{\prime}$. In these expressions we have introduced the function
\begin{align}
W _{m} (x) = \epsilon _{1} \mbox{K} _{m} (x) \mbox{I} _{m} ^{\prime} (x) - \epsilon _{2} \mbox{K} _{m} ^{\prime} (x) \mbox{I} _{m} (x) .
\end{align}
Of course, if we set $\epsilon _{1} = \epsilon _{2} = 1$, we recover the vacuum result, $g _{m} (\rho , \rho ^{\prime}) = \mbox{I} _{m} (k \rho _{<}) \, \mbox{K} _{m} (k \rho _{>})$, as we must \cite{Schwinger}.

Now, we repeat this calculation for $\rho ^{\prime} < a$ (that is, the source
lies in the medium 1). The final outcome for the reduced free GF, in the medium 1 ($\rho < a$), is
\begin{align}
\mathfrak{g} _{m} (\rho , \rho ^{\prime}) &= \frac{1}{\epsilon _{1}} \mbox{I} _{m} (k \rho _{<}) \, \mbox{K} _{m} (k \rho _{>}) + \mbox{I} _{m} (k \rho) \, \mbox{I} _{m} (k \rho ^{\prime}) \notag \\ & \phantom{=} \times \frac{\mbox{K} _{m} (k a)}{\mbox{I} _{m} (k a)} \left[ \frac{1}{ka} \frac{1}{W _{m} (k a)} - \frac{1}{\epsilon _{1}} \right] , \label{gfrho'<a1}
\end{align}
and, in the medium 2 ($\rho > a$), 
\begin{align}
\mathfrak{g} _{m} (\rho , \rho ^{\prime}) = \frac{1}{ka} \frac{\mbox{I} _{m} (k \rho ^{\prime}) \mbox{K} _{m} (k \rho )}{W _{m} (k a)} . \label{gfrho'<a2}
\end{align}

\end{document}